\documentclass[prd,aps,a4paper,eqsecnum,twocolumn,nofootinbib,floatfix]{revtex4}  %

\newif\ifusesec
\usesectrue  
   
\usepackage{graphicx} 
\usepackage{amsmath,amsfonts,amssymb}
\usepackage{mathtools}
\usepackage{color}

\newcommand{\be}{\begin{equation}}
\newcommand{\ee}{\end{equation}}
\newcommand{\g}{\gamma}
\newcommand{\al}{\alpha}
\newcommand{\pinf}{p_{\infty}}
\newcommand{\eq}{\eqref}
\newcommand{\bh}{\bar h}
\newcommand{\bn}{\bar \nu}
\newcommand{\nn}{\nonumber \\}

\newcommand{\beq}{\begin{equation}}
\newcommand{\eeq}{\end{equation}}
\newcommand{\bea}{\begin{eqnarray}}
\newcommand{\eea}{\end{eqnarray}}

\begin{document}

 \title{Conservative binary dynamics beyond order $\alpha^5$ in electrodynamics}

\author{Donato Bini$^{1}$, Thibault Damour$^2$}
  \affiliation{
$^1$Istituto per le Applicazioni del Calcolo ``M. Picone,'' CNR, I-00185 Rome, Italy\\
$^2$Institut des Hautes Etudes Scientifiques, 91440 Bures-sur-Yvette, France
}
\date{\today}

\date{\today}

\begin{abstract}
We compute the conservative scattering angle of two classical charged particles at the sixth order in electromagnetic coupling,
and at the fourth order in velocity, thereby going beyond the current state of the art [fifth order in coupling, derived 
by Bern {\it et al.}, Phys. Rev. Lett. \textbf{132}, 251601 (2024)]. Our result is obtained by using the 
electromagnetic version of the Effective One-Body formalism to transfer information from the exact circular binary-charge 
solution of Schild [Phys. Rev. \textbf{131}, 2762 (1963)] to the post-Lorentzian expansion of the scattering angle.
 \end{abstract}

\maketitle

\section{Introduction}

The detection of gravitational-wave signals \cite{LIGOScientific:2016aoc} has motivated a renewed effort to  derive ever more
accurate analytical descriptions of relativistic corrections to the motion of binary systems. In particular,
Refs. \cite{Damour:2016gwp,Damour:2017zjx} pointed out that the post-Minkowskian (PM)
approximation method (expansion in powers of the gravitational coupling constant, $G$, taking into account all powers of
$\frac{v}{c}$) can be usefully combined with the Effective One-Body (EOB) approach \cite{Buonanno:1998gg,Buonanno:2000ef,Damour:2001tu} to gravitational-wave template construction. Recent years have witnessed spectacular progresses in PM gravity
computations, notably by leveraging advances in Quantum-Field-Theory (QFT) scattering, e.g.  \cite{Bern:2019crd,Bern:2021dqo,Herrmann:2021tct,Bjerrum-Bohr:2021din,Bini:2021gat,Bern:2021yeh,Dlapa:2021vgp,Dlapa:2022lmu,Bini:2022enm,Damgaard:2023ttc,Driesse:2024xad}.
These advances in PM gravity have, in turn, motivated studies of the technically simpler post-Lorentzian (PL) approximation method
(expansion in powers of the electromagnetic coupling constant, $\alpha$, taking into account all powers of
$\frac{v}{c}$), see Refs. \cite{Saketh:2021sri,Bern:2021xze,Wang:2022ntx,Bern:2023ccb,Ceresole:2023wxg,Cheung:2024zdq}.

The current state of the art in (classical) PL electrodynamics is the {\it conservative} 5PL level, i.e. the $O(\al^5)$
knowledge of the electromagnetic scattering angle of two charges interacting in a time-symmetric way derived
by Bern {\it et al.} \cite{Bern:2023ccb}.
The (classical)  time-symmetric electromagnetic interaction of charges was pioneered by Fokker \cite{Fokker:1929} and 
pursued by Wheeler and Feynman \cite{Wheeler:1945ps,Wheeler:1949hn}. From the QFT viewpoint, it corresponds to keeping
only the potential-photon contributions to the classical scattering angle. Radiation-reaction contributions to the scattering angle
have been considered at order $O(\al^3)$ in Refs. \cite{Saketh:2021sri,Bern:2021xze,Wang:2022ntx}. Ref. \cite{Bern:2023ccb} further considered
the  $O(\al^4)$ ``conservative radiation contribution" to the scattering angle, linked to the square of the
radiation-reaction force (see also Appendix C of \cite{Porto:2024cwd}).

The aim of the present work is to use the exact (time-symmetric) solution of the circular motion of two charges due to
 Schild \cite{Schild:1963} to derive information about the conservative electromagnetic scattering beyond
 $O(\al^5)$, and notably at PL orders $O(\al^6)$ and $O(\al^7)$. The crucial tool we shall use to 
 transfer information between the particular (circular) bound-state solution of Schild,
 and scattering-state solutions, is the electromagnetic version of the EOB approach \cite{Ceresole:2023wxg,Buonanno:2000qq}.
 Specifically, we encode (as detailed in the next Section) the conservative electrodynamics in an energy-dependent Coulomb-gauge EOB potential, $\Phi(E,r)$,
 admitting a PL expansion of the general form
 \beq \label{phiPL0}
\Phi(E, r)=\sum_{n \geq1} \al^n \frac{\Phi_n(E)}{r^n}\,,
\eeq
where $E$ denotes the total energy of the system in the center-of-mass (cm) frame.

We use the EOB, energy-gauge potential $\Phi(E,r)$ as a go-between relating  two
gauge-invariant functions, $\chi(E,J)$ and $H(J, I_r)$, respectively characterizing scattering states
and bound states. Here, $\chi(E,J)$ is the scattering function (with $J$ denoting the total cm angular momentum), 
while  $H(J, I_r^{\rm ell})$ (with $I_r^{\rm ell}$ denoting the bound-state radial action) is the Delaunay Hamiltonian
(which reduces to $E_{\rm circ}(J)$ along circular motions). The present work considers general charged binaries,
with arbitrary charges, $e_1$ and $e_2$, and arbitrary masses, $m_1$ and $m_2$.

\section{EOB description of the two-body conservative electromagnetic interaction}

We consider an EOB description of the {\it conservative} dynamics of two charges interacting
in a Fokker-Wheeler-Feynman way. One starts from the mass-shell condition of a particle of mass $\mu= \frac{m_1m_2}{(m_1+m_2)}$,  effective energy $-P_0= {\cal E}$, and total angular momentum $P_\phi=J$,
of the form \cite{Ceresole:2023wxg}
\be
\eta^{\mu\nu} (P_\mu - e_{\rm eff}{ A}_\mu) (P_\nu - e_{\rm eff} {A}_\nu) + \mu^2=0\,.
\ee
We recall that while the EOB angular momentum $P_\phi=J$ is equal to the real  angular momentum of the two-body
system, the effective energy ${\cal E}$ is related to the real  energy $E$ of the binary system by the EOB energy map:
\be
{\cal E}= \frac{E^2 - m_1^2 -m_2^2}{2 (m_1+m_2) }\,,
\ee
or, equivalently,
\be
\frac{E}{M}=\sqrt{1+ 2 \nu \left(\frac{{\cal E}}{\mu} -1 \right) }\,,
\ee
 with the notation, $M \equiv m_1+m_2$, and
\be
\nu \equiv \frac{\mu}{M}=\frac{m_1 m_2}{(m_1+m_2)^2}\,.
\ee
We use an EOB  Coulomb gauge where ${A}^i=0$ and $  e_{\rm eff}{A}^0={\cal V}(\g, r)$, leading to
\bea
0 &=& - [{\cal E}- {\cal V}(\g, r)]^2 + \mu^2 + {\bf P}^2\nonumber\\
&=&  - [{\cal E}- {\cal V}(\g, r)]^2 + \mu^2 + P_r^2+ \frac{J^2}{r^2}\,.
\eea

In the probe limit,  $\frac{m_1}{m_2} \to 0$, or $\nu \to 0$ (we assume here $m_1<m_2$), the EOB description should reduce to the dynamics of a probe charge $e_1$
in the (Coulomb) potential of a heavy charge $e_2$, so that we have (in Gauss units)
\be \label{probelimit}
\lim_{\nu \to 0} {\cal V}(\g, r)= \frac{e_1 e_2}{r}\,.
\ee

It is convenient to rescale everything by the effective mass $\mu$ with
\be
p_\mu= \frac{P_\mu}{\mu} \,;\, \g = \frac{{\cal E}}{\mu}   \,;\, p_r= \frac{P_r}{\mu} \,;\, j=  \frac{P_\phi}{\mu}= \frac{J}{\mu} \,;\, \phi = \frac{{\cal V}}{\mu}\,. 
\ee
The $\mu^2$-rescaled mass-shell condition reads
\be
0= - (\g - \phi)^2 + 1 + p_r^2+ \frac{j^2}{r^2}\,,
\ee
or
\be
p_r^2 =  (\g - \phi)^2-1 - \frac{j^2}{r^2}\,.
\ee
Note that we have not rescaled the radius $r$ which has units of length.
In the above equation $p_r^2$, $\g$ , and $\phi$ are dimensionless (using $c=1$), while $j$ has units of length.
We often use
\be
u \equiv \frac1r\,,
\ee
and introduce the effective potential
\bea
{\mathcal W}(j, \g, u)&\equiv&  (\g - \phi)^2-1 -j^2 u^2 \nonumber\\
&=&\g^2-1 - 2\gamma \phi(\g, u) + \phi^2(\g, u)- j^2 u^2\,,\nonumber\\
\eea
so that the mass-shell condition yields
\be
p_r^2 = {\mathcal W}(j, \g, u) \,.
\ee
This mass-shell condition, and therefore the (energy-dependent) effective potential ${\mathcal W}(j, \g, u)$, encapsulates the 
full information about the electromagnetic binary dynamics, both for scattering states (when $\g>1$) and for bound states (when $\g<1$).
The aim of this work is to transfer information between the current perturbative knowledge of scattering states \cite{Bern:2023ccb}, 
and the exact description of classical {\it circular} bound states \cite{Schild:1963}. 
The EOB, energy-dependent, potential $\phi(\g,\nu,u)$ plays the role of a go-between relating scattering states and bound states.
The transfer of information between scattering states and bound states, via the potential $\phi(\g,\nu,u)$, goes through some 
gauge-invariant functions (called ``radial actions") characterizing either scattering states (hyperbolic radial action),  
or bound states (elliptic radial action), as described in the following Sections.

\section{Relation between the EOB potential and the hyperbolic radial action (and the scattering function) }

A basic property of the scattering-EOB approach is that the effective scattering angle defined by the
EOB dynamics coincides with the real scattering angle of the two-body system \cite{Damour:2016gwp}.
This allows one to relate the physical scattering function, $\chi_{\rm real}(E,J)$, to the one
obtained by computing the EOB scattering angle, $\chi$, induced by the potential $\phi$. 

The latter  is obtained from the integral
\be
\pi+\chi=-\int_{- \infty}^{+ \infty} dr  \frac{\partial}{\partial J} P_r = -\int_{- \infty}^{+ \infty} dr  \frac{\partial}{\partial j} p_r\,, 
\ee
i.e.,
\beq \label{chi1}
\pi+\chi= + 2j \int_0^{u_{\rm max}}\frac{du}{\sqrt{{\mathcal W}(j, \g, u)}}\,,
\eeq
where $u_{\rm max} = \frac1{r_{\rm min}}$ is the inverse radial turning point. 
Note in passing that the integral expression \eq{chi1} for the scattering angle, here obtained in Coulomb gauge, 
$ e_{\rm eff} {A}^{\rm C}_\mu(r) \propto (-\phi^{\rm C}(r), {\mathbf 0})$,
coincides with the corresponding expression for $\chi$ obtained in Ref. \cite{Ceresole:2023wxg} which used
a Kerr-Schild-type gauge, with $e_{\rm eff} A^{\rm KS}_\mu \propto (-\phi^{\rm KS}(r), + \phi^{\rm KS}(r) \frac{\bf r}{r} )$ . The reason behind this agreement  between two different parametrizations of $ e_{\rm eff}{A}_\mu$ 
(namely $\phi^{\rm C}(r) \equiv \phi^{\rm KS}(r)$)  is that the use
of  an electromagnetic gauge-transformation $ d \Lambda(r) \propto \phi(r) dr$ gauges away  the radial part of $e_{\rm eff} {A}^{\rm KS}_\mu $.

The radial integral appearing on the right-hand side (rhs) of Eq. \eq{chi1} is convergent. It is usually considered 
in the perturbative regime of large impact parameters, where the electromagnetic EOB potential $\phi$  is represented as
a post-Lorentzian (PL) expansion of the type
\beq \label{phiPL}
\phi(\g, \nu, u)=\sum_{n \geq1}  \frac{\phi_n(\g, \nu)}{r^n}= \sum_{n \geq1} \phi_n(\g, \nu) u^n\,.
\eeq
Note that $\phi$ being dimensionless, $ \phi_n$ is a  length$^n$. The PL expansion of $\phi$, Eq. \eq{phiPL}, is
the electromagnetic analog of the post-Minkowskian (PM) expansion of the corresponding EOB description of the gravitational
dynamics of binary systems. Essentially, PL expansions are expansions in powers of the electromagnetic coupling constant $e_1 e_2 \propto \alpha$,
while PM expansions are expansions in powers of Newton's constant $G$.

Inserting the PL expansion of the potential, Eq.  \eq{phiPL}, in Eq. \eq{chi1}, leads (after computing the resulting integrals
either by Cauchy contour integrals \cite{sommerfeld}, or by Hadamard-regulated real integrals \cite{Damour:1988mr})
 to the following PL expansion of $\chi$ in powers of $\frac{e_1 e_2}{J}$, i.e. in powers of $\frac1j$:
\beq
\chi(\g,j)=\sum_{n \geq1} \frac{2 \, \chi_n(\g)}{j^n}\,.
\eeq
Here again $\chi$ being dimensionless, $ \chi_n$ is a  length$^n$.

The hyperbolic radial action is formally defined as a regularized version of the (divergent) integral, 
$I_r^{\rm hyp}(E,J) =\int_{- \infty}^{+ \infty} dr  P_r$, i.e. a function of $E$ and $J$ (or, equivalently, $\g$ and $j$) such that
\be
\pi+ \chi(E,J) =-\frac{\partial}{\partial J} I_r^{\rm hyp}(E,J) = - \frac1{\mu} \frac{\partial}{\partial j} I_r^{\rm hyp}(\g,j)\,.
\ee
The PL expansion of $ I_r^{\rm hyp}(\g,j)/\mu$ reads
\be
\frac{ I_r^{\rm hyp}(\g,j)}{\mu}= - \pi j - 2 \chi_1(\g) \ln j+ \sum_{n\geq 2} \frac{2 \chi_n(\g)}{(n-1) j^{n-1}}\,.
\ee
Inserting the  the PL expansion of the potential,  \eq{phiPL}, in Eq. \eq{chi1}, leads to an explicit link between
the $\chi_n$'s and the $\phi_n$'s. The first seven orders read 
\begin{widetext} 
\bea \label{chinvsphin}
\chi_1 &=&  -\frac{\phi_1 \gamma}{\sqrt{\gamma^2 - 1}}\,,\nonumber\\
\chi_2 &=& \frac{\pi}{4} (\phi_1^2 -2\gamma \phi_2  )\,,  \nonumber\\
\chi_3 &=& -\frac{\gamma(2\gamma^2-3)}{3(\gamma^2-1)^{3/2}}\phi_1^3
+\frac{2 (2\gamma^2-1)}{(\gamma^2-1)^{1/2}}\phi_1\phi_2
-2\gamma (\gamma^2-1)^{1/2}  \phi_3\,,\nonumber\\
\chi_4 &=&\pi\left( \frac{3}{16}\phi_1^4-\frac94 \gamma\phi_1^2\phi_2+\frac34 (3\gamma^2-1)\phi_1\phi_3 +\frac38 (3\gamma^2-1)\phi_2^2 -\frac34 \gamma (\gamma^2-1)\phi_4\right)\,, \nonumber\\
\chi_5 &=& -\frac{\gamma (8\gamma^4-20\gamma^2+15)}{15(\gamma^2-1)^{5/2}} \phi_1^5
+\frac{4 (8\gamma^4-12\gamma^2+3)}{3(\gamma^2-1)^{3/2}}\phi_1^3\phi_2
-\frac{4 (4\gamma^2-3)\gamma}{(\gamma^2-1)^{1/2}}\phi_1^2\phi_3
-\frac{4(4\gamma^2-3)\gamma}{(\gamma^2-1)^{1/2}}\phi_2^2\phi_1\nonumber\\
&+&\frac{8(\gamma^2-1)^{1/2}(4\gamma^2-1)}{ 3} \phi_4 \phi_1
+\frac{8 (\gamma^2-1)^{1/2}(4\gamma^2-1)}{3}\phi_2 \phi_3
-\frac{8\gamma (\gamma^2-1)^{3/2}}{3}\phi_5\,, \nonumber\\
\chi_6 &=&\pi\left(\frac{5}{32}\phi_1^6
-\frac{75}{16}\gamma \phi_1^4  \phi_2
+\frac{15}{8}(5\gamma^2-1)\phi_3 \phi_1^3
+ \frac{45}{16}(5\gamma^2-1)\phi_2^2\phi_1^2  
-\frac{15}{8}\gamma(5\gamma^2-3)\phi_4\phi_1^2\right.\nonumber\\
&-&  \frac{15}{4}\gamma(5\gamma^2-3)\phi_3\phi_2 \phi_1
-\frac{5}{8}\gamma (5\gamma^2-3)\phi_2^3
+\frac{15}{16}(\gamma^2-1) (5\gamma^2-1)\phi_4\phi_2 \nonumber\\
&+&\left. \frac{15}{32}(\gamma^2-1) (5\gamma^2-1)\phi_3^2
+\frac{15}{16}(\gamma^2-1)(5\gamma^2-1)\phi_5\phi_1
-\frac{15}{16} \gamma (\gamma^2-1)^2\phi_6  
\right)\,, \nonumber\\
\chi_7 &=& -\frac{\gamma (70\gamma^2+16\gamma^6-35-56\gamma^4)}{35(\gamma^2-1)^{7/2}} \phi_1^7
+\frac{6(16\gamma^6-40\gamma^4+30\gamma^2-5)}{5(\gamma^2-1)^{5/2}} \phi_1^5 \phi_2
-\frac{2 \gamma (24\gamma^4-40\gamma^2+15)}{(\gamma^2-1)^{3/2}}  \phi_1^4\phi_3\nonumber\\
& -&\frac{4 (24\gamma^4-40\gamma^2+15)\gamma}{(\gamma^2-1)^{3/2}} \phi_2^2\phi_1^3
+\frac{8 (8\gamma^4-8\gamma^2+1)}{(\gamma^2-1)^{1/2}} \phi_4\phi_1^3
+ \frac{24 (8\gamma^4-8\gamma^2+1)}{ (\gamma^2-1)^{1/2}}\phi_2\phi_3\phi_1^2\nonumber\\
&-&24 \gamma (2\gamma^2-1)(\gamma^2-1)^{1/2} \phi_5\phi_1^2
+\frac{8 (8\gamma^4-8\gamma^2+1)}{(\gamma^2-1)^{1/2}} \phi_2^3 \phi_1
-48 \gamma (2 \gamma^2-1)(\gamma^2-1)^{1/2}\phi_2\phi_4\phi_1\nonumber\\
&-&24 \gamma (2\gamma^2-1)(\gamma^2-1)^{1/2}\phi_3^2 \phi_1
-24 \gamma (2\gamma^2-1)(\gamma^2-1)^{1/2}  \phi_2^2\phi_3
+\frac{16}{5} (\gamma^2-1)^{3/2}  (6\gamma^2-1) \phi_3\phi_4
\nonumber\\
&+&\frac{16}{5} (\gamma^2-1)^{3/2}  (6\gamma^2-1) \phi_2\phi_5
+\frac{16 (6\gamma^2-1) (\gamma^2-1)^{3/2}}{5}  \phi_1\phi_6
-\frac{16}{5} (\gamma^2-1)^{5/2} \gamma  \phi_7\,.
\eea
\end{widetext}
Some features of the functional dependence of the scattering coefficients $\chi_n$'s on the potential coefficients 
$\phi_n$'s displayed in Eq. \eq{chinvsphin} can be noted: (i) the $n$-th order coefficient $\chi_n$ only depends on $\phi_1, \phi_2,\cdots, \phi_n$; and (ii) the dependence of $\chi_n$ on $\phi_n$ involves a coefficient $\propto \pinf^{n-2}$, while
the dependence of $\chi_n$ on $\phi_{n-1}$ involves a coefficient $\propto \pinf^{n-4}$.  

Similarly to what happens in gravity \cite{Damour:2019lcq},  the PL-expansion 
(for a given impact parameter $b$) of the four-velocity variation, $\Delta u_a^\mu$, of each body ($a=1,2$) during scattering
possesses a joint polynomiality structure in the masses, $m_a$, and the mass-ratios, $\hat e_a \equiv \frac{e_a}{m_a}$
(see \cite{Cheung:2024zdq}). In the present, conservative case, this polynomiality structure is best seen by introducing the  electromagnetic analog of Newton's constant, $G_{\rm EM}= \hat e_1 \hat e_2=\frac{e_1 e_2}{m_1 m_2}$, and by noticing
that $(G_{\rm EM})^n$ factors out at the $n$th PL order. One then finds (following the reasoning of \cite{Damour:2019lcq})  that the rescaled scattering coefficients
\be
\tilde \chi_n(\g,\nu) \equiv h^{n-1}  \chi_n(\g,\nu) \,;\qquad h \equiv \sqrt{1+2\nu(\gamma-1)}\,,
\ee
have the following structure
\bea \label{tildechistructure}
\tilde \chi_n(\g,\nu)&=& A^n \left[ \hat \chi_{n,0}(\g) + \nu  \hat \chi_{n,1}(\g) +\cdots\right.\nonumber\\
&+&\left. \nu^{d(n)} \hat\chi_{n,d(n)}(\g) \right]\,,
\eea
where we factored out the $n$th power of
\be
A\equiv \frac{ e_1 e_2}{\mu}\,,
\ee
(which has  the dimension of a length, so that the $\hat \chi_{n,k}(\g)$ are dimensionless),
where $\nu \equiv \frac{m_1 m_2}{(m_1+m_2)^2}$  is the symmetric mass ratio, and where the order of the
polynomial in $\nu$ describing $ \tilde \chi_n(\g,\nu)$ is $d(n) \equiv \left[ \frac{n-2}{2} \right]$,
with $\left[ \cdots \right]$ denoting the integer part. [Note that $A$ is negative when considering bound states
of opposite charges.]  The polynomial order $d(n)$ takes the following values (of relevance below): 
\bea
&&d(1)=d(2)=0 \; ; \; d(3)=d(4)=1\; ; \nonumber\\
&&d(5)=d(6)=2 \; ; \; d(7)=d(8)=3\; ; \cdots
\eea
The full implications of the special structure, \eq{tildechistructure}, of  $\tilde \chi_n(\g,\nu)$ for the structure of
the $\phi_n(\g,\nu)$ will be discussed below. Let us only remark here that, using the probe-limit property \eqref{probelimit},
which implies
\be
\lim_{\nu \to 0}\phi(\g, \nu, u)= \frac{e_1 e_2}{\mu\,  r} = \frac{A}{r}\,,
\ee
we deduce that: $\phi_1(\g, \nu=0)= A$, and for $n\geq 2$,  $\phi_n(\g, \nu=0)=  0$. The $\nu$-polynomial structure, \eq{tildechistructure}, together with the link, $ \chi_1 =  -\frac{\phi_1 \gamma}{\sqrt{\gamma^2 - 1}}$, then implies
that $\frac{\phi_1(\g, \nu)}{A}$ must be independent of $\nu$ and therefore be simply equal to
\be
\phi_1(\g, \nu)=A\,.
\ee

\section{Relation between the EOB potential and the elliptic radial action (and the Delaunay Hamiltonian) }

Bound states are classically encoded in two  (equivalent) gauge-invariant functions: the elliptic radial action
\be
I_r^{\rm ell}(E,J) \equiv  \oint dr  P_r =   2 \mu  \int_{r_{\rm min}}^{r_{\rm max}} dr \sqrt{{\mathcal W}\left(j, \g, \frac1r\right)}\,,
\ee
which is defined here without dividing it by $2 \pi$, or the Delaunay Hamiltonian,
\be
E= H(I_r, J)\,,
\ee
obtained by solving the equation $I_r= I_r^{\rm ell}(E,J)$ with respect to $E$.
The variation of the polar angle from periastron to periastron (including its basic $2 \pi$ contribution) is related to the radial 
action via
\be
\Phi(E,J)= -\frac{\partial}{\partial J} I_r^{\rm ell}(E,J) \,.
\ee

The gauge-invariant nature of these functions played a useful role in the study of the gravitational interaction
of binary systems \cite{Damour:1988mr}, and in the inception of the EOB approach \cite{Buonanno:1998gg}.

Ref. \cite{Kalin:2019inp} pointed out the existence of a simple relation (valid for  potentials having a quasi-local nature) between the periastron-advance function,
$\Phi(E,J)$, and the scattering function, $\chi(E,J)$, namely
\be
\Phi(E,J)={\rm AC}\left[ (\pi+\chi(E,J)) + (\pi +\chi(E,-J)) \right]\,,
\ee
where the notation AC refers to a procedure of analytic continuation in the two variables $E$ and $J$ ($E$ varying
from the scattering domain, $E> m_1+m_2; \g>1$, to the bound-state one, $E < m_1+m_2; \g<1$, and $J$ varying from
$+J$ to $-J$). It was noticed in \cite{Bini:2020nsb} that the latter analytic continuation  naturally extends to the 
corresponding radial actions in the form
\be
I_r^{\rm ell}(E,J) = {\rm AC}\left[ I_r^{\rm hyp}(E,J) -  I_r^{\rm hyp}(E,-J)\right]\,,
\ee
thereby yielding  
\be \label{Irell1}
\frac{ I_r^{\rm ell}(\g,j)}{\mu}= - 2\pi j   - 2\pi \frac{A \gamma}{\sqrt{1-\gamma^2 }}+ \sum_{k\geq 1} \frac{4 \chi_{2k}(\g)}{(2k-1) j^{2k-1}}\,.
\ee
Here we used the expression of $\chi_1$ in terms of $\phi_1$ (the analytic continuation of $\ln j$ around $j=0$, together
with that of $\frac{ \gamma}{\sqrt{\gamma^2 - 1}}$ around $\g=1$, yield the second term on the rhs), and the fact that
the even coefficients $\chi_{2k}(\g)$ have no branch cuts at $\g=1$. 
By inserting in the elliptic-motion results \eq{Irell1}, the relations \eq{chinvsphin} between the $\chi_n$'s and the
the $\phi_n$'s, we get the link between the dynamics of bound states  and the PL expansion of the EOB potential $\phi(\g,r)$.
The latter link, expressing the coefficient of $\frac1{j^{2k-1}}$ in $ I_r^{\rm ell}(\g,j)$ as a polynomial in the 
potential coefficients $\phi_n$'s, can also be directly obtained by expanding in a post-Coulombian (PC) way (i.e. a combined
expansion in $ v^2 \sim \frac{e_1 e_2}{\mu r}$) the radial integral $\oint dr  P_r $ defining the  elliptic radial action.

We note in passing (following \cite{Bini:2020nsb}) that the special structure \eq{tildechistructure} of $\tilde \chi_n$
 implies a corresponding $\nu$-polynomiality structure of the elliptic radial action, namely
\be \label{Irell2}
\frac{ I_r^{\rm ell}(\g,j)}{\mu}= - 2\pi j   - 2\pi \frac{A \gamma}{\sqrt{1-\gamma^2 }}+ \sum_{k\geq 1} \frac{4 A^{2k} \hat \chi_{2k}(\g,\nu)}{(2k-1)(h j)^{2k-1}}\,,
\ee
where $\hat \chi_{2k}(\g,\nu)$ is a polynomial of order $k-1$ in $\nu$.

Here, we shall focus on {\it circular} (bound) orbits. They correspond to considering the limit $ I_r^{\rm ell}(\g,j) \to 0$ in the above
equations. Solving the equation $ I_r^{\rm ell}(\g,j) = 0$ yields the dependence of the energy on the angular momentum along the
sequence of circular orbits:
\be \label{Ecirc}
E= E_{\rm circ}(J)=H(I_r=0, J)\,.
\ee
Other observable functions of the circular dynamics can be defined by using as argument the orbital frequency $\omega$, e.g. $E_{\rm circ}(\omega)$
 or $J_{\rm circ}(\omega)$. However, these functions do not contain more information because of the identity,
 $\omega=dE_{\rm circ}(J)/dJ$.

The link between the energetics of circular motions, i.e. the function $E= E_{\rm circ}(J)$, and the PL expansion of the EOB potential 
$\phi(\g,r)$ can be more directly obtained by using the effective potential $ {\mathcal W}(j,\g, u)$. Namely, by solving
the two equations
\bea
 {\mathcal W}(j,\g, u) &=&0\,, \nonumber\\
 \frac{\partial}{\partial u} {\mathcal W}(j,\g, u)&=& 0\,.
\eea
The second equation can be solved in $j^2$, namely (the prime denoting a $u$ derivative)
\be \label{ju}
j^2=-\frac{(\gamma-\phi(u))\phi'(u)}{u}\,. 
\ee
Given $\phi(u)$ this yields a parametric representation of $j^2$ in terms of $u$. Inserting Eq. \eqref{ju}
in the equation $  {\mathcal W}(j,\g, u)=0$  then yields
\be \label{gu}
(\g-\phi)^2 + (\g-\phi) u \phi'= 1\,.
\ee
Formally eliminating $u$ between Eqs. \eq{ju} and \eq{gu} gives the link between $\g$ and $j$,
or, equivalently between the real energy $E$ and the real angular momentum $J = \mu j$, i.e. the
circular-energy function \eq{Ecirc}, or, equivalently, its inverse: $J= J_{\rm circ}(E)$.
Note that this yields a relation between two physical observables.
By contrast, the parametric representations \eq{ju}, and \eq{gu}, involve the EOB radial coordinate $r=\frac1u$,
which has no direct observable meaning, because it is obtained from the real distance between the two
bodies by applying a canonical transformation.
 
One can perturbatively solve Eqs. \eq{ju} and \eq{gu} in the large-circular-orbit limit where $u \to 0$, $j^2 \to \infty$ and
 $\g -1 \to 0$. The leading order (Coulomb) solution yields $ u \approx - \frac{A}{j^2}  $
 and $\g -1 \approx -\frac12  \frac{A^2}{j^2}$.
Inserting in Eqs. \eq{ju} and \eq{gu} the expansion $\phi= A u + A^2 \hat \phi_2 u^2 +  A^3 \hat \phi_3 u^3 + \cdots$, with $A <0$,
one can perturbatively eliminate $u$ between these two equations and derive various PC-expanded links between 
$\g-1$ and $\frac{A^2}{j^2}$. 
For instance, one can write $\g-1$ as a power series in $\frac{A^2}{j^2}$ of the form
\bea
\label{gvsy}
\g-1&=&  -\frac12  \frac{A^2}{j^2}+ \left(  \hat\phi_2 -\frac18  \right)  \frac{A^4}{j^4} \nonumber\\
&+&  \left(  \hat\phi_2 - 2 {\hat\phi}_2^2 - \hat \phi_3 - \frac1{16}  \right)  \frac{A^6}{j^6}\nonumber\\
&+& \left(- \frac{5}{128} +\hat\phi_2- 5\hat\phi_2^2 + 4 \hat\phi_2^3 -\frac32  \hat\phi_3\right.\nonumber\\ 
&+&\left. 6\hat\phi_2  \hat\phi_3  + \hat\phi_4  \right)     \frac{A^8}{j^8}\nonumber\\
&+& O\left(   \frac{A^{10}}{j^{10}} \right)\,.
\eea
 This PC-expanded relation is recursive in the sense that the potential coefficients $\hat \phi_n$ appearing on the rhs are functions of $\g$
 (and therefore depend on the formal PC expansion parameter $\g-1 \sim \frac{A^2}{j^2}$). We have checked that the $\frac{A^2}{j^2}$-expanded
 link, Eq. \eq{gvsy}, here
 obtained directly from the circular conditions, Eqs. \eq{ju} and \eq{gu}, agrees with the various PC-expanded links
 between $\frac{A}{j}$ and  $\g-1$ which can be derived by  recursively solving in $j$ 
 the  equation $ I_r^{\rm ell}(\g,j) = 0$, which reads (from Eq. \eq{Irell1})
 \be \label{jvschi2k}
 j = -  \frac{A \gamma}{\sqrt{1-\gamma^2 }}+ \sum_{k\geq 1} \frac{2 \chi_{2k}[\phi_1,\ldots,\phi_{2k}]}{\pi (2k-1) j^{2k-1}}\,.
 \ee
One should note that in the expansion \eq{jvschi2k}, the {\it fractional} $k$-PC term 
($\propto  \frac{ 1}{ j^{2k}}$)  enters with a coefficient $\propto \chi_{2k}$, which a priori involves $\hat \phi_2, \hat \phi_3, \cdots$
up to $ \hat \phi_{2k}$. However, the high-order potential terms $ \hat \phi_{2k}$, $ \hat \phi_{2k-1}$, etc. appear
with PC-small coefficients, namely $\pinf^{2k -2}$, $\pinf^{2k -4}$, etc. As a consequence, only  
$\hat \phi_2, \hat \phi_3, \cdots$,  up to $\hat \phi_{k+1} $ contribute at the  fractional $k$-PC level (as seen
in the few terms displayed in Eq. \eq{gvsy}).

\section{Gauge-invariant description of the probe limit}

To have some idea of the structure of the gauge-invariant functions, notably $\chi(E,J)$ and $E_{\rm circ}(J)$, 
respectively describing scattering motions, and circular motions, it is useful to pause to study the probe limit $\nu \to 0$.
As already said, in this limit the EOB potential $\phi(\g,\nu,u)$ simply reduces to the ($\mu$-rescaled) Coulomb potential  
\be
\phi_{\nu \to 0} = \frac{A}{r}= A \, u\,,
\ee
where we recall that $A\equiv \frac{ e_1 e_2}{\mu}$.

It is easy to evaluate the scattering angle in the probe limit, with the result, e.g. \cite{Bern:2023ccb},
\bea \label{chiprobe}
\frac12\chi^{\nu \to 0} &=& \frac{1}{ \sqrt{1 - \frac{A^2}{j^2}} } \left( \frac{\pi}{2}-\arctan \left[  \frac{\g}{\sqrt{\g^2-1} } \frac{\frac {A}{j}}{ \sqrt{1 - \frac{A^2}{j^2}}}\right]\right)\nonumber\\
&-& \frac{\pi}{2}\,,  
\eea
here written in a form valid for both signs of  $\frac {A}{j}$ near zero.

The  PL expansion of $\frac12\chi^{\nu \to 0}$ (i.e its expansion in powers of $\frac Aj$) reads
\bea  \label{chiprobeexp}
\frac12  \chi^{\nu \to 0} &=& - \frac{\g}{\sqrt{\g^2-1} } \frac {A}{j} + \frac{\pi}{4}  \frac {A^2}{j^2} 
- \frac{ \gamma (-3 + 2 \gamma^2) }{ 3 (\gamma^2-1)^{3/2} }\frac{A^3 }{j^3} \nonumber\\
&+& \frac{3  \pi}{16 }\frac{A^4}{j^4} 
- \frac{\gamma  (15 - 20 \gamma^2 + 8 \gamma^4)}{15(\gamma^2-1)^{5/2} }\frac{A^5 }{j^5} 
+ \frac{5 \pi}{32 } \frac{A^6 }{j^6}\nonumber\\
&+&  \frac{ \gamma  (35 - 70 \gamma^2 + 56 \gamma^4 - 16 \gamma^6) }{(35(\gamma^2-1)^{7/2} } \frac{ A^7}{j^7}
+ \frac{35 A^8\pi}{256 j^8}\nonumber\\
&+&O\left(\frac{A^9}{j^9}\right)\,. 
\eea
Concerning circular motions (with $A <0$), it is easy, in the probe limit, say $m_1 \ll m_2$, to eliminate $u$ between Eqs. \eq{ju} and \eq{gu}, with the following result
for the relation between the specific binding energy\footnote{In the probe limit, the total energy $E$ tends to $m_2$.} $\g =\lim_{m_1\to 0} \frac{E-m_2}{m_1}$
and the rescaled angular momentum $j = \frac{J}{\mu}$
\be \label{gvsjprobe}
\g_{\rm circ}^{\nu \to 0}(j) = \sqrt{1 - \frac{A^2}{j^2}}\,.
\ee
In the large-circular-orbit region, $j \to \infty$, $\g_{\rm circ}$ admits an expansion in inverse powers of $\frac{A^2}{j^2}$,
namely
\bea
\gamma|_{\rm circ}^{\nu=0}(j)&=&1 - \frac12 \frac{A^2}{j^2}   -\frac18 \frac{A^4}{j^4}  - \frac{1}{16} \frac{A^6}{j^6} 
  - \frac{5}{128}\frac{A^8}{j^8}  \nonumber\\
&-&  \frac{7}{256 }\frac{A^{10}}{j^{10}} - \frac{21}{1024}    
 \frac{A^{12}}{j^{12}} -\frac{33}{2048} \frac{A^{14}}{j^{14}}\nonumber\\ 
&-&   \frac{429}{32768}\frac{A^{16}}{j^{16}} 
  - \frac{715}{65536}\frac{A^{18}}{j^{18}}\nonumber\\ 
&+& O\left(\frac{A^{20}}{j^{20}}\right)\,.
\eea

In the other limit of small circular orbits, the sequence of circular orbits exists down to zero radii, $r \to 0$, $u\to \infty$,
with the notable fact that, in this limit, the velocity $v_1$ of $m_1$ tends to the velocity of light,
$v_1 \to 1$, while the angular momentum tends to a finite minimum, namely $ j_{\rm min}=  |A|$,
and the specific binding energy tends to zero in a square-root manner: 
$\g_{\rm circ}^{\nu \to 0}(j) = \sqrt{1 - \frac{ j_{\rm min}^2}{j^2}}$.

\section{Structure of the EOB-potential coefficients $\phi_n(\g,\nu)$ inherited from the polynomiality structure of the 
scattering-angle coefficients $\chi_n(\g,\nu)$. }

By combining the relations \eq{chinvsphin} between the 
scattering-angle coefficients $\chi_n(\g,\nu)$, and the EOB-potential coefficients $\phi_n(\g,\nu)$, with the
 information \eq{tildechistructure} about the $\nu$-structure of the $\chi_n(\g,\nu)$'s, one can get a rather
 precise description of the structure of the $\phi_n(\g,\nu)$'s as functions of $h= \sqrt{1+ 2 \nu(\g-1)}$ and $\nu$.
 A crucial fact allowing us to find the structure of the $\phi_n(\g,\nu)$'s is that the coefficients entering the relations
 \eq{chinvsphin} depend only on $\g$ but not on $\nu$ (or $h$). The analog investigation in PM gravity was pursued
 in Section X of \cite{Bini:2020nsb} (see the discussion there of the EOB gravity potential coefficients $q_n(\g,\nu)$).
 
 Some notation is convenient for deriving the structure of the $\phi_n(\g,\nu)$'s. First, similarly to the
 definition $\tilde \chi_n \equiv h^{n-1}\chi_n$, let us define
 \be
  \tilde \phi_n \equiv h^{n-1}\phi_n\,.
 \ee
 Second, it is useful to consider the dependence of both $\tilde \chi_n$  and $\tilde \phi_n$ on the formal quadratic extension
 of the real values of $\g$ obtained by adding the square-root $h= \sqrt{1+ 2 \nu(\g-1)}$ (considered as an analog of $ i = \sqrt{-1}$).
To start with, we note that it is easily seen, by multiplying both sides of the relations  \eq{chinvsphin} by $h^{n-1}$, and by 
using the polynomiality result \eq{tildechistructure}, that $\tilde \phi_n$ is a polynomial of order $n-1$ in $h$, with coefficients
depending only on $\g$. In addition, when $n \geq2$, $\tilde \phi_n$ must vanish in the probe limit, i.e when $ h \to 1$. It is then
convenient to introduce the following additional notation
\be
\bar h \equiv h-1\; ;\; \bar \nu \equiv 2 \nu (\g-1)\;.
\ee
With this notation, the basic quadratic identity satisfied by $h= \sqrt{1+ 2 \nu(\g-1)}=  \sqrt{1+ \bar \nu}$, reads
\be \label{hid}
\bar h^2 = - 2 \bar h + \bar \nu\,.
\ee
Recursively using the quadratic identity \eq{hid} allows one to transform any polynomial in $h$, or equivalently in $\bh$, to a 
{\it first order} polynomial in $\bh$, with coefficients which are polynomials in $\bn$, e.g. $\bh^3= (4+\bn) \bh - 2 \bn$.
[This is analogous to transforming any polynomial $P(z)$ in a complex variable $z = x+ i y$ to the form $P_1(x,y) + i P_2(x,y)$
with real  polynomials $P_1$ and $P_2$.]

Let us rewrite the $\nu$ structure of $\tilde \chi_n$, Eq.\eq{tildechistructure}, as
\beq  \label{tildechistructure2}
\tilde \chi_n(\g,\nu)=\tilde \chi_{n,0}(\g)+\tilde \chi_{n, {\bar 1}}(\g)\bar \nu+\ldots +\tilde \chi_{n,{\bar d(n)}}(\g)\bar \nu^{d(n)}
\eeq
where the zeroth-order term $\tilde \chi_{n,0}$ is simply given by the probe limit,
\be
\tilde \chi_{n,0}(\g)= A^n \hat \chi_{n,0}(\g)\,,
\ee
as given by the PL expansion of Eq. \eq{chiprobe}, i.e. by Eq. \eq{chiprobeexp}.
To distinguish the coefficients $\tilde \chi_{n, {\bar k}}$ in Eq. \eq{tildechistructure2} from the ones, $\tilde \chi_{n, k}$, entering Eq. \eq{tildechistructure} (they differ by a factor $[2(\g-1)]^k$),
we add a bar over the second index $k$.
In other words, the genuine post-probe information about the structure of the scattering coefficients is
entirely contained in the coefficients $\tilde \chi_{n, {\bar k}}(\g)$ with $n \geq1$.

Recursively using Eq. \eq{hid} one can easily prove (by induction) that the $\phi_n$'s have the following structure:
\beq
\tilde \phi_n(\g, \nu)=\bar h \, \tilde \phi^{(1)}_{n}(\g, \bar \nu)+
\tilde \phi^{(2)}_{n}(\g, \bar \nu)\,,
\eeq
where
\bea
\tilde \phi^{(1)}_{n}(\g, \bar \nu)&=& \tilde \phi^{(1)}_{n,0}(\g)+\tilde \phi^{(1)}_{n, \bar 1}(\g)\bar \nu+\cdots\nonumber\\ 
&+& \tilde \phi^{(1)}_{n, \bar d(n-1)}(\g)\bar \nu^{d(n-1)}\,,\nonumber\\
\tilde \phi^{(2)}_{n}(\g, \bar \nu)&=&\tilde \phi^{(2)}_{n, \bar 1}(\g)\bar \nu+\ldots \tilde \phi^{(2)}_{n,\bar d(n)}(\g)\bar \nu^{d(n)}\,, 
\eea
with the important knowledge that, at any PL order $n$, the coefficients $\tilde \phi^{(1)}_{n,\bar k}(\g)$ entering the $\bh$-linear
term depend only on the  $\tilde \chi_{n',\bar k'}(\g)$'s belong to the previous orders (i.e. $n'<n$), while the  $\tilde \chi_{n, \bar k}(\g)$'s
belonging to order $n$ only enter the coefficients $\tilde \phi^{(2)}_{n,\bar k}(\g)$ of the $\bh$-independent term.

More explicitly we have
\bea
\tilde \phi_2 &=& \bar h \, \tilde \phi^{(1)}_{2,0} \nonumber\\
\tilde \phi_3 &=& \bar h \, \tilde \phi^{(1)}_{3,0}+ \tilde \phi^{(2)}_{3,\bar 1}\bar \nu \nonumber\\
\tilde \phi_4 &=& \bar h \,[\tilde \phi^{(1)}_{4,0}+\tilde \phi^{(1)}_{4,\bar 1}\bar \nu]+ \tilde \phi^{(2)}_{4,\bar 1}\bar \nu \nonumber\\
\tilde \phi_5 &=& \bar h \,[\tilde \phi^{(1)}_{5,0}+\tilde \phi^{(1)}_{5,\bar 1}\bar \nu]+ \tilde \phi^{(2)}_{5,\bar 1}\bar \nu+\tilde \phi^{(1)}_{5,\bar 2}\bar \nu^2\nonumber\\
\tilde \phi_6 &=& \bar h \,[\tilde \phi^{(1)}_{6,0}+\tilde \phi^{(1)}_{6,\bar 1}\bar \nu+\tilde \phi^{(1)}_{6,\bar 2}\bar \nu^2]+ \tilde \phi^{(2)}_{6,\bar 1}\bar \nu+\tilde \phi^{(2)}_{6,\bar 2}\bar \nu^2\nonumber\\
\tilde \phi_7 &=& \bar h \,[\tilde \phi^{(1)}_{7,0}+\tilde \phi^{(1)}_{7,\bar 1}\bar \nu+\tilde \phi^{(1)}_{7,\bar 2}\bar \nu^2]\nonumber\\
&+& \tilde \phi^{(2)}_{7,\bar 1}\bar \nu+\tilde \phi^{(2)}_{7,\bar 2}\bar \nu^2+\tilde \phi^{(2)}_{7,\bar 3}\bar \nu^3\,.\nonumber\\
\eea
Similarly to what happens in the gravitational case \cite{Damour:2019lcq}, a remarkable fact happens at order $n=2$. Namely, the 
2PL scattering coefficient is fully determined by the probe limit, namely
\be
\chi_2(\g,\nu)= \frac{\chi_{2,0}(\g)}{h}=  \frac{\pi}{4 } \frac{A^2}{h}\,,
\ee
so that the 2PL potential coefficient $\tilde \phi_2$ is also determined by the probe limit:
\be
 \tilde \phi^{(1)}_{2,0}=\frac{A^2}{2\gamma}\,, \qquad {\rm i.e.} \qquad \phi_2= \frac{A^2 \, \bh}{ 2 \g h}\,.
 \ee
 It is only starting at the 3PL order that the genuine new information contained in the $\tilde \chi_{n,\bar k}(\g)$'s starts 
 entering the potential coefficients. Specifically, we find, for $n=3,4,5,6$ (recalling that in these expressions, the coefficients
  $\tilde \phi^{(1,2)}_{n, \bar k}(\g)$'s are to be multiplied by $\bn^k$ and not simply by $\nu^k$):
  
\begin{widetext}
\bea \label{phi3vschi}
\frac{\tilde \phi^{(1)}_{3,0}}{A^3} &=& -\frac{ 2\gamma^2 - 1 }{2\gamma^2 (\gamma^2 - 1)}\,,\nonumber\\
\frac{\tilde \phi^{(2)}_{3,\bar 1}}{A^3} &=&  -\frac{1}{2\gamma (\gamma^2 - 1)^{1/2}}\tilde \chi_{3,\bar 1} 
+ \frac{ 4\gamma^4 - 6\gamma^2 + 3 }{6\gamma^2 (\gamma^2 - 1)^2}\,,
\eea

\bea \label{phi4vschi}
\frac{\tilde \phi^{(1)}_{4,0}}{A^4} &=& \frac{ 10\gamma^4 - 9\gamma^2 + 3}{4\gamma^3 (\gamma^2 - 1)^2 }\,,\nonumber\\
\frac{\tilde \phi^{(1)}_{4,\bar 1}}{A^4} &=&  -\frac{(3\gamma^2 - 1)}{2\gamma^2(\gamma^2-1)^{3/2}}\tilde \chi_{3,\bar 1}  
+ \frac{ 27\gamma^6 - 49\gamma^4 + 45\gamma^2 - 15 }{24\gamma^3(\gamma^2-1)^3}\,,\nonumber\\
\frac{\tilde \phi^{(2)}_{4,\bar 1}}{A^4} &=&   -\frac{3\gamma^2 - 1)}{2\gamma^2(\gamma^2-1)^{3/2} }\tilde \chi_{3,\bar 1} 
- \frac{4}{ 3\gamma \pi (\gamma^2-1)}\tilde \chi_{4,\bar 1}  
 - \frac{ 27\gamma^6 - 53\gamma^4 + 21\gamma^2 - 3 }{ 24\gamma^3 (\gamma^2 - 1)^3}\,, 
\eea

\bea \label{phi5vschi}
\frac{\tilde \phi^{(1)}_{5,0}}{A^5} &=& \frac{ -28\gamma^6 + 35\gamma^4 - 20\gamma^2 + 5 }{ 4\gamma^4(\gamma^2-1)^3 }\,, \nonumber\\
\frac{\tilde \phi^{(1)}_{5,\bar 1}}{A^5}  &=&\frac{  (4\gamma^2 - 1)}{4\gamma^3(\gamma^2-1)^{3/2}} \tilde \chi_{3,\bar 1}
-\frac{4 (4\gamma^2 - 1)}{ 3\gamma^2\pi (\gamma^2-1)^2 }\tilde \chi_{4,\bar 1}
-\frac{(64\gamma^6 - 76\gamma^4 + 54\gamma^2 - 15)}{ 12\gamma^4(\gamma^2-1)^{3}}\,,\nonumber\\
\frac{\tilde \phi^{(2)}_{5,\bar 1}}{A^5}  &=&   -\frac{(12\gamma^4 - 5\gamma^2 + 2)}{ 4\gamma^3(\gamma^2-1)^{5/2}}\tilde \chi_{3,\bar 1} 
 - \frac{4 (4\gamma^2 - 1)}{3\gamma^2\pi(\gamma^2-1)^2}\tilde \chi_{4,\bar 1} 
 - \frac{3}{8\gamma(\gamma^2-1)^{3/2}}\tilde \chi_{5,\bar 1}  
+ \frac{(432\gamma^8 - 950\gamma^6 + 965\gamma^4 - 480\gamma^2 + 105)}{ 120\gamma^4(\gamma^2-1)^4}\,, \nonumber\\
\frac{\tilde \phi^{(2)}_{5,\bar 2}}{A^5}  &=&
 -\frac{(16\gamma^4 - 10\gamma^2 + 3)}{4\gamma^3(\gamma^2-1)^{5/2}}\tilde \chi_{3,\bar 1}  
- \frac{3}{8\gamma(\gamma^2-1)^{3/2}}\tilde \chi_{5,\bar 2}  
+ \frac{(256\gamma^8 - 520\gamma^6 + 660\gamma^4 - 420\gamma^2 + 105)}{120\gamma^4 (\gamma^2-1)^4}\,, 
\eea

\bea \label{phi6vschi}
\frac{\tilde \phi^{(1)}_{6,0}}{A^6} &=& \frac{ (2\gamma^2 - 1) (42\gamma^6 - 45\gamma^4 + 28\gamma^2 - 9)}{ 4\gamma^5 (\gamma^2-1)^4 } \,, \nonumber\\
\frac{\tilde \phi^{(1)}_{6,\bar 1}}{A^6} &=&  -  \frac{(45\gamma^6 - 42\gamma^4 + 22\gamma^2 - 5)}{ 4\gamma^4(\gamma^2-1)^{7/2}}\tilde \chi_{3,\bar 1} 
+\frac{ 2 (5\gamma^2 - 1) }{ 3\gamma^3(\gamma^2-1)^2\pi }\tilde \chi_{4,\bar 1} 
- \frac{(3(5\gamma^2 - 1))}{ 8 \gamma^2(\gamma^2-1)^{5/2}}\tilde \chi_{5,\bar 1} \nonumber\\
&+& \frac{(5625\gamma^{10} - 13899\gamma^8 + 16250\gamma^6 - 11710\gamma^4 + 4845\gamma^2 - 855)}{ 240\gamma^5(\gamma^2-1)^5)}\,, \nonumber\\
\frac{\tilde \phi^{(1)}_{6,\bar 2}}{A^6} &=&  \frac{(5\gamma^2 - 1)}{8\gamma^3(\gamma^2-1)^2)}\tilde \chi_{3,\bar 1}^2 
-\frac{ (125\gamma^6 - 109\gamma^4 + 63\gamma^2 - 15)}{ 12\gamma^4 (\gamma^2-1)^{7/2} } \tilde \chi_{3,\bar 1}
-\frac{(3  (5\gamma^2 - 1))}{8\gamma^2 (\gamma^2-1)^{5/2} }\tilde \chi_{5,\bar 2}\nonumber\\ 
&+& \frac{(3125\gamma^{10} - 6841\gamma^8 + 10710\gamma^6 - 9750\gamma^4 + 4725\gamma^2 - 945)}{ 720\gamma^5(\gamma^2-1)^5}\,,\nonumber\\
\frac{\tilde \phi^{(2)}_{6,\bar 1}}{A^6} &=&  -\frac{20\gamma^6 - 5\gamma^4 + 7\gamma^2 - 2}{4\gamma^4(\gamma^2-1)^{7/2}}\tilde \chi_{3,\bar 1}
 - \frac{4 (10\gamma^4 - 3\gamma^2 + 1)}{3\gamma^3 (\gamma^2-1)^3\pi)} \tilde \chi_{4,\bar 1}
- \frac{(3 (5\gamma^2 - 1))}{8(\gamma^2-1)^{5/2} \gamma^2)} \tilde \chi_{5,\bar 1}
- \frac{16}{15\gamma\pi (\gamma^2-1)^2} \tilde \chi_{6,\bar 1}\nonumber\\
&-& \frac{(1250\gamma^{10} - 3253\gamma^8 + 3260\gamma^6 - 1970\gamma^4 + 690\gamma^2 - 105)}{120\gamma^5(\gamma^2-1)^5} \,,\nonumber\\
\frac{\tilde \phi^{(2)}_{6,\bar 2}}{A^6} &=&   \frac{(5\gamma^2 - 1)}{8\gamma^3(\gamma^2-1)^2}\tilde \chi_{3,\bar 1}^2 
- \frac{((25\gamma^6 + \gamma^4 + 9\gamma^2 - 3) }{ 6\gamma^4(\gamma^2-1)^{7/2}  }\tilde \chi_{3,\bar 1} 
- \frac{2 (25\gamma^4 - 12\gamma^2 + 3)}{ 3\gamma^3(\gamma^2-1)^3\pi)}\tilde \chi_{4,\bar 1} 
- \frac{ 3 (5\gamma^2 - 1) }{ 8\gamma^2 (\gamma^2-1)^{5/2}}\tilde \chi_{5,\bar 2} \nonumber\\
&-& \frac{16}{ 15\gamma\pi (\gamma^2-1)^2} \tilde \chi_{6,\bar 2}
- \frac{3125\gamma^{10} - 7927\gamma^8 + 8010\gamma^6 - 5430\gamma^4 + 2025\gamma^2 - 315}{360\gamma^5(\gamma^2-1)^5}\,. 
\eea
\end{widetext}
Another structural information to keep in mind concerning these expressions, is that while the odd-PL-order scattering coefficients
$\chi_{2 m+1}$ are singular in the zero-velocity limit ($\g \to 1$), namely $\chi_{2 m+1} \sim (\g^2-1)^{- (2m+1)/2}$ (as in the probe limit),
the even-PL-order ones are regular when $\g \to 1$ (and contain an overall factor $\pi$). By contrast, the potential coefficients
$\phi_n$ are all regular when $\g \to 1$. However, the separate pieces $\bar h \,\tilde \phi^{(1)}_{n}$ and $\tilde \phi^{(2)}_{n}$
 are often singular in the zero-velocity limit; their various $(\g^2-1)^{-m/2}$ singularities compensating each other when $\g \to 1$.
[The overall factors $\pi$ in the even-order $\chi_n$'s also disappear in the $\phi_n$'s.]

\section{Explicit value of the EOB-potential $\phi(\g,\nu,u)$ through accuracy $O(\alpha^5)$}

Bern et al.  \cite{Bern:2023ccb} (see Eqs. (9) and (17) there) computed the $O(\alpha^5)$-accurate conservative scattering 
angle.  For illustration, let us cite here the explicit  $O(\alpha^4)$-accurate results of Ref. \cite{Bern:2023ccb} :
\bea
\chi_1 &=& -\frac{A\gamma}{\sqrt{\gamma^2 - 1}}\,,\nonumber\\
\chi_2 &=&  \frac{A^2\pi}{4 h}\,,\nonumber\\
\chi_3 &=& \frac{A^3(-2\gamma (2\gamma^2 - 3) + 4\nu (\gamma  - 1) (\gamma^3 + 3\gamma^2 - 3))}{6 h^2 (\gamma^2 - 1)^{3/2}},  \nonumber\\
\chi_4 &=&  \frac{A^4 3\pi }{8h^3}\left[ 1+\nu\left(C_4^{\rm c}(\gamma)+C_4^{\ln}(\gamma) L + C_4^{\ln^2}\!(\gamma) L^2\right)\right]\,, \nn 
\eea
where $L \equiv \ln x \equiv \ln(\gamma-\sqrt{\gamma^2-1})$ and 
\bea
C_4^{\rm c}(\gamma)&=&\frac{1}{2(\gamma^2-1)}\left( 3\gamma^4-11 \gamma^3+3\gamma^2+\gamma+14\right.\nonumber\\
&-&\left.\frac{7\gamma^2-1}{\gamma^3}\right)\,, \nonumber\\
C_4^{\ln{}}(\gamma)&=&\frac{(3\gamma^3-4\gamma^2+9\gamma-4)}{ (\gamma^2-1)^{3/2}}\,,  \nonumber\\
C_4^{\ln^2{}}(\gamma)&=& \frac{3\gamma^2+1}{2(\gamma^2-1)^2}\,.
\eea
Above  the  variable $x$,  defined as,
 \be
  x \equiv \gamma- \sqrt{\gamma^2-1} \,;\quad \hbox{with}\quad 0<x<1\,,
 \ee
 rationalizes both $\g$ and $\pinf$:
\beq
\gamma=\frac12 \left(\frac1 x + x \right)\,,\quad \pinf \equiv \sqrt{\gamma^2-1}=\frac12\left(\frac{1}{x}-x \right)\,.
\eeq
Inserting the  $O(\alpha^5)$-accurate scattering angle of  \cite{Bern:2023ccb} in our
Eqs. \eq{phi3vschi},\eq{phi4vschi},\eq{phi5vschi} above, yields the corresponding
$O(\alpha^5)$-accurate value of the EOB potential $\phi$. Let us display here the explicit expressions
of $\phi_1$,  $\phi_2$,  $\phi_3$,  and  $\phi_4$. The explicit expression of  $\phi_5$ is too involved to be displayed here and is
given in the ancillary file to the arxiv submission of this paper.
\bea \label{phi123}
\frac{\phi_1}{A}&=& 1\,, \nonumber\\
\frac{\phi_2}{A^2}&=&  \frac{1}{2\g} \left(1-\frac1{h}  \right) \,,\nonumber\\
\frac{\phi_3}{A^3}&=&\frac{2\gamma^2-1}{2\gamma^2(\gamma^2-1)}\left(1-\frac{1}{h}\right)\nonumber\\
&-&\frac{1}{2\gamma (\gamma-1)}\left(1-\frac{1}{h^2}\right), \nn
\frac{\phi_4}{A^4}&=&B_4^{(1)}\left(1-\frac{1}{h}\right)+B_4^{(2)}\left(1-\frac{1}{h^2}\right)\nonumber\\
&+&B_4^{(3)}\left(1-\frac{1}{h^3}\right)\,,
\eea
where
\bea
B_4^{(1)}&=& B_4^{\ln{}}+\frac{1}{16\gamma^4 (\gamma^2-1)^2 (\gamma-1)}(1-12\gamma +5\gamma^2\nonumber\\
&+&  46\gamma^3+31\gamma^4-33\gamma^5+25\gamma^6+3\gamma^7)\nonumber\\
B_4^{(2)}&=& -\frac{(3\gamma^2-1)(5\gamma^2-1)}{8\gamma^3 (\gamma^2-1)(\gamma-1)}\nonumber\\
B_4^{(3)}&=& -B_4^{\ln{}}-\frac{1}{16\gamma^4 (\gamma^2-1)^2 (\gamma-1)}(1-7\gamma^2 \nonumber\\
&+&10\gamma^3 +5\gamma^4+7\gamma^5-15\gamma^6+3\gamma^7)\nonumber\\
B_4^{\ln{}}&=& \frac{(3\gamma^2-1)}{16\gamma (\gamma^2-1)^3(\gamma-1)}L^2\nonumber\\
&+& \frac{-4+9\gamma-4\gamma^2+3\gamma^3}{8\gamma (\gamma^2-1)^{5/2}(\gamma-1)}L\,.\nonumber\\
\eea
For the reason explained above, the values of $\phi_1$, $ \phi_2$ and  $ \phi_3$ derived here coincide with the
corresponding results of \cite{Ceresole:2023wxg} (when discarding the radiative contributions to $\phi_3$ in
Eq. (20) there, i.e. the contributions odd in $\pinf$).

Here, we displayed for more visibility the expressions of the $\phi_n$'s without decomposing them in
their $\bh$-linear parts and their $\bh$-independent  (pure $\nu$-dependent) parts. Let us only
indicate the $\bh$-linear part, and the pure $\nu$-dependent part of $\tilde \phi_3=h^{2}\phi_3$:
\be
\frac{\tilde \phi_3}{A^3}= \frac{h^{2}\phi_3}{A^3}=\hat \phi_3^{\bh} +  \hat \phi_3^{\bn}\,,
\ee
with
\bea
\hat \phi_3^{\bh} &=&- \bh \frac{ 2\gamma^2 - 1 }{2\gamma^2 (\gamma^2 - 1)}\,, \nonumber\\
\hat \phi_3^{\bn} &=& \bn \frac{\g^2-\g-1}{2\gamma^2(\gamma^2-1)} \,.
\eea

Let us also give the post-Coulomb (PC) expansions (in terms of $p_\infty \equiv\sqrt{\gamma^2-1}$) of the
potential coefficients at the total PC order that we have worked, namely the overall 6PC order, when taking
into account that each order in $A$ counts also as one PC order (using the virial-theorem scaling, namely $A \sim  \pinf^2$)
\bea \label{phinPCexp}
\frac{\phi_1^{\rm PC}}{A} &=&  1 \,,\nonumber\\
\frac{\phi_2^{\rm PC}}{A^2} &=&  \frac14  \nu p_\infty^2 - \frac{3}{16}  \nu (1 + \nu) p_\infty^4 + 
 \frac{1}{32} \nu (5 + 6 \nu + 5 \nu^2) p_\infty^6\nonumber\\ 
&-& 
 \frac{5}{256} \nu (7 + 9 \nu + 10 \nu^2 + 7 \nu^3) p_\infty^8 \nonumber\\
&+& 
 \frac{21}{512} \nu (3 + 4 \nu + 5 \nu^2 + 5 \nu^3 + 3 \nu^4)p_\infty^{10}\nonumber\\ 
&+& O(p_\infty^{12})\,,\nonumber\\
\frac{\phi_3^{\rm PC}}{A^3} &=& -\frac34  \nu  + \frac{1}{16}   \nu (11 + 13 \nu) p_\infty^2 - \frac{3}{32}   \nu (7 + 9 \nu + 9 \nu^2)  p_\infty^4\nonumber\\
&+& 
 \frac{1}{256} \nu (163 + 217 \nu + 266 \nu^2 + 221 \nu^3) p_\infty^6\nonumber\\ 
&-& 
 \frac{1}{512} \nu (319 + 433 \nu + 575 \nu^2 + 640 \nu^3 + 449 \nu^4)p_\infty^8\nonumber\\ 
&+& O(p_\infty^{10}) \,,\nonumber\\
\frac{\phi_4^{\rm PC}}{A^4} &=&  -\frac{1}{72}  \nu (79 + 45 \nu)\nonumber\\ 
&+&  
 \frac{1}{1440}\nu (2062 + 2235 \nu + 1620 \nu^2) p_\infty^2 \nonumber\\
&-&  \frac{\nu (116639 + 135765 \nu + 144550 \nu^2 + 104475 \nu^3)}{67200} p_\infty^4\nonumber\\ 
&+&  \frac{1}{2419200}\nu (4837654 + 5973741 \nu + 6923700 \nu^2\nonumber\\ 
&+& 7020825 \nu^3 + 4687200\nu^4)p_\infty^6\nonumber\\  
&+& O(p_\infty^8)\,,\nonumber\\
\frac{\phi_5^{\rm PC}}{A^5} &=& -\frac{1}{144}  \nu (215 + 143 \nu + 63 \nu^2)\nonumber\\ 
&+&  
 \frac{\nu (41457 + 43505 \nu + 27775 \nu^2 + 
    15975 \nu^3)}{14400}p_\infty^2 \nonumber\\  
&-&\frac{1}{14112000} \nu (62916073 + 83428975 \nu +  68359900 \nu^2 \nonumber\\
&+& 43603875 \nu^3 + 27286875 \nu^4)  p_\infty^4\nonumber\\  
&+& O(p_\infty^6)\,.
 \eea

In the following Section we shall use the old result of Schild \cite{Schild:1963} to go beyond the current $O(\alpha^5)$ knowledge
 and to obtain information about the  $O(\alpha^6)$ and  $O(\alpha^7)$ levels, namely about the coefficients
 entering
\bea \label{phi67PCexp}
\frac{ \phi_6^{\rm PC}}{A^6} &=& \hat \phi_6^{(0)} +\hat \phi_6^{(2)} p_\infty^2 + O(p_\infty^4)\,,\nonumber\\
\frac {\phi_7^{\rm PC}}{A^7} &=& \hat \phi_7^{(0)} +  O(p_\infty^2)\,.
\eea
The present work mainly relies on the low-velocity behavior of the potential $\phi(\g,\nu,u)$. Let us remark in
passing that, in the high-energy limit, $\g \to \infty$, or $x=\frac1{\g+\pinf} \approx \frac1{2 \g} \to 0$, the potential
$\phi(\g,\nu,u)$ tends to its probe limit: $\lim_{\g \to \infty} \phi(\g,\nu,u)= \phi_1 =A \, u$. More precisely, we have, when $x \to 0$:
\bea
\phi_2&=& A^2 x+O(x^{3/2})\,,\nonumber\\
\phi_3&=& 2A^3x^2 +O(x^{5/2})\,,\nonumber\\
\phi_4&=&  -\frac{3}{4\sqrt{\nu}} A^4 x^{5/2}+O(x^3)\,,\nonumber\\
\phi_5&=& -\frac{6}{\sqrt{\nu}}  A^5 x^{7/2}+O(x^{9/2}) \,.
\eea
The property that $\lim_{\g \to \infty} \phi(\g,\nu,u)= \phi_1 =A \, u$ was also found to hold (through $O(\al^3)$)
for the radiation-reacted version of $\phi(\g,\nu,u)$ in Ref. \cite{Ceresole:2023wxg}.

\section{Schild's exact solution for two massive charges in circular motion}

Consider a system of two   masses  $m_1$, $m_2$ (with total mass   $M \equiv m_1+m_2$ and  reduced mass
$\mu = m_1 m_2/(m_1+m_2)$), and equal (and opposite) charges ($e_1=-e_2=e$) in circular motion with velocities $v_1$ and $v_2$. Schild \cite{Schild:1963} has found an exact solution of the associated Fokker-Wheeler-Feynman (time-symmetric) dynamics as described by the following relations (involving the auxiliary angle $\theta$ measuring the retardation effect in the 
electromagnetic interaction, and the orbital circular frequency $\omega$)
\bea
\label{theta}
&& v_1^2 + v_2^2 + 2v_1v_2\cos\theta - \theta^2= 0\,, 
\eea
and
\begin{widetext}
\bea
\label{eq1}
&&\frac{m_1 v_1}{\omega \sqrt{1-v_1^2}}- \frac{e^2}{(\theta + v_1v_2\sin\theta)^3} [(v_1 + v_2\cos\theta) (1 - v_1^2) (1 - v_2^2) + (v_1\theta + v_2\sin\theta) (\theta + 
v_1 v_2\sin\theta) ]=0\,, \nn
&&\frac{m_2v_2}{\omega \sqrt{1-v_2^2}}
 - \frac{e^2}{(\theta + v_1v_2\sin\theta)^3}[(v_2 + v_1\cos \theta) (1-v_1^2) (1-v_2^2) + (v_2\theta  + v_1\sin\theta) (\theta + v_1v_2\sin\theta)]=0\,.
\eea
Eliminating $\omega$ between the last two equations yields the compatibility relation
\bea \label{v}
\frac{m_1 v_1 \sqrt{1-v_2^2}}{m_2v_2 \sqrt{1-v_1^2}}&=& \frac{ (v_1 + v_2\cos\theta) (1 - v_1^2) (1 - v_2^2) + (v_1\theta + v_2\sin\theta) (\theta + 
v_1 v_2\sin\theta)  }{ (v_2 + v_1\cos \theta) (1-v_1^2) (1-v_2^2) + (v_2\theta  + v_1\sin\theta) (\theta + v_1v_2\sin\theta) }\,.
\eea
\end{widetext}
Schild also derived the conserved energy and angular momentum of the system (involving interaction terms), namely
\bea
\label{E}
E&=& m_1\sqrt{1 - v_1^2} + m_2\sqrt{1 - v_2^2}\,, \\
\label{J}
J&=& e^2 \frac{ 1 + v_1v_2\cos\theta }{ \theta + v_1v_2\sin \theta }\,. 
\eea
As before, we find convenient to parametrize  the total energy $E$ of the system in terms of  the effective energy $\gamma$, with
\bea
\label{eff_energy}
E&=&Mh=M\sqrt{1+2\nu(\gamma-1)}\,,\nonumber\\ 
\gamma&=&1 + \frac{1}{2\nu} \left(\frac{E^2}{M^2} - 1\right)\,.
\eea 

Because of the mixture of algebraic and trigonometric dependence on $\theta$ one cannot algebraically solve
the two equations \eq{theta}, \eq{v} to get $v_1$ and $v_2$ as explicit functions of $\theta$.
However, one can numerically solve Eqs. \eq{theta}, \eq{v} to compute the variation of $v_1$ and $v_2$ with $\theta$.
Inserting these results in Eqs. \eq{E} and \eq{J}, then allows one to numerically compute the parametric variation of $E$ and $J$
in terms of $\theta$. 

For any finite value of the mass ratio, say $q \equiv \frac{m_2}{m_1} \geq 1$, 
the  domain of variation of $\theta$ is the real interval $[0, \theta_m]$, where $\theta_m$ is the positive solution of
the equation $ \theta^2=2(1+\cos\theta) $, or, equivalently
\be
\frac{\theta}{2}=\cos \frac{\theta}{2}\,.
\ee
Its numerical value is
\be
\theta_m= 1.478170266430\ldots
\ee
The limit $\theta \to 0^+$ corresponds to $v_1 \to 0$ and $v_2 \to 0$, i.e. to large orbits, large values of $J$ and small
binding energies, $E/M \to 1$, $\g \to 1$. The other limit $\theta \to \theta_m^-$ corresponds to $v_1 \to 1^-$ and $v_2 \to 1^-$,
and to orbits with vanishingly small radii (and $\omega \to \infty$). As $\theta$ runs over the interval $[0, \theta_m]$ the
energy $E$ monotonically decreases from $M$ down to zero, while the angular momentum $J$  monotonically decreases from 
$+ \infty$ down to a finite value $J_m$ reached as $\theta \to \theta_m^-$. [This contrasts with the GR case, where the
sequence of circular orbits passes through a maximally bound orbit at a finite radius (``last stable circular orbit")
before running through unstable circular orbits, with increasing energies, as the radius decreases down to the ``light-ring" radius.]
The minimum value $J_m$ of $J$ reached for vanishingly small orbits is given by (for any finite mass ratio $q$)
\beq \label{Jm}
{\mathcal J}_m \equiv \frac{{J}_m}{e^2}=\frac{1+\cos\theta_m}{\theta_m+\sin \theta_m}= 0.4416107917053\ldots
\eeq

Of particular interest is the behavior of the energy at the end of the sequence of circular orbits (i.e for vanishingly small radii).
One finds the following square-root behavior (denoting ${\mathcal J} \equiv \frac{{J}}{e^2}$)
\beq  \label{sqrtvanishing}
\frac{E}{M} \approx \sqrt{1-2\nu}  \frac{ 4{\mathcal J}_m}{1+{\mathcal J}_{m}^2}\sqrt{\frac{{\mathcal J}}{{\mathcal J}_{\rm min}}-1}\,,
\eeq
where the factor in front of the square-root is equal to
\be
 \frac{ 4{\mathcal J}_m}{1+{\mathcal J}_{m}^2}= \theta_m\,.
\ee
Note in passing that the formal probe limit, $m_1/m_2 \to 0$, is exceptional in that the sequence of circular orbits ends
(for vanishingly small radii) when $v_1 \to 1$ with $v_2 =0$, so that the minimum value of ${\mathcal J}$ is not
equal to ${\mathcal J}_m=  0.4416107917053\ldots$, but rather to  the larger value ${\mathcal J}_m^{\rm probe}=1$
(corresponding to ${ J}_m^{\rm probe}=e^2$, i.e. $j_m = \frac{e^2}{\mu} = - A = |A|$).
As already indicated in Eq. \eq{gvsjprobe}, the energy has, at the end of the sequence of circular orbits,  also a square-root behavior in 
the probe limit, though this can only be seen when considering the specific binding energy $(E-m_2)/m_1 \approx \g$. [The ratio
$E/M$ tends to 1 in the probe limit.]

\begin{figure}
\includegraphics[scale=0.30] {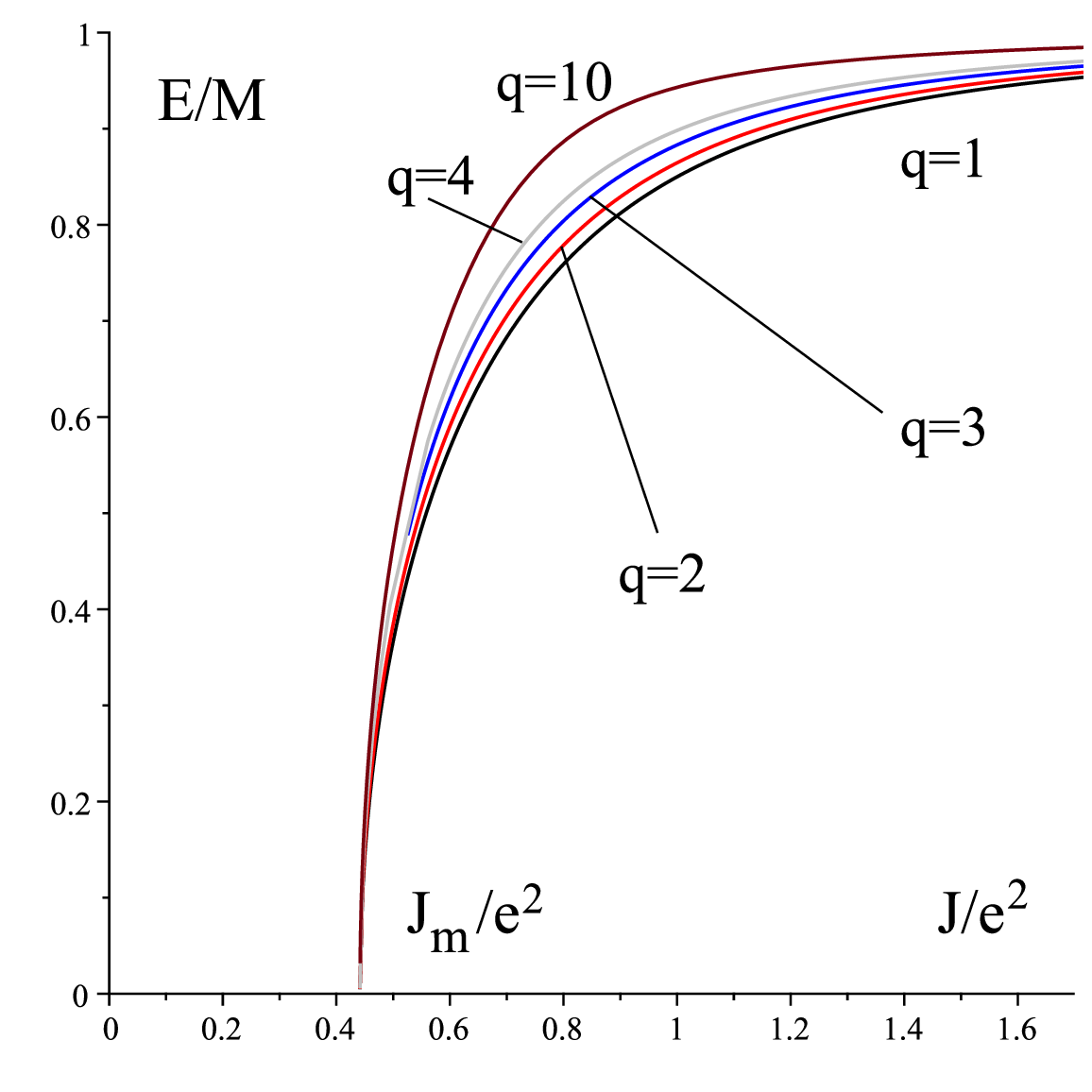}
\caption{\label{figEvsJ} The behavior of $E/M$ versus $J/e^2$ is shown for different values of $q=\frac{m_2}{m_1}=[1,2,3,4,10]$. The minimum value  $J_m/e^2 \approx  0.4416107917053\ldots$ is independent of $q$. See text for details.}
\end{figure}

Fig. \ref{figEvsJ} illustrates the variation of the total energy $E/M$ as a function of ${\mathcal J}=J/e^2$ along the
sequence of circular orbits, for various values of the mass ratio $q = \frac{m_2}{m_1} \geq 1$.

The limit of large, weakly-bound, circular orbits (with $ J \to \infty$) can be treated analytically. 
Indeed, working in a PC approximation, i.e. in the small-velocity regime $v_1 \sim v_2 \sim \theta \ll 1$, one can perturbatively solve
the two equations Eqs. \eq{theta}, \eq{v} to get $v_1$ and $v_2$ as power series in  $\theta$. The beginning of these power series
is 
\bea
v_1(\theta)&=& \frac{m_2}{M} \theta + \frac{m_1m_2^2}{2M^3} \theta^3\nonumber\\ 
&+&\frac{ m_1 m_2 (12 m_1^3 + 11 m_1^2 m_2 - 5 m_1 m_2^2 - 13 m_2^3) }{24M^5}\theta^5\nonumber\\
&+& O(\theta^7)\,,
\eea
with the expansion of $v_2(\theta)$ obtained by swapping $m_1 \leftrightarrow m_2$.

Inserting these expansions in Eqs. \eq{E} and \eq{J} then gives the expansions of $E/M$ (or $\g$) and $J/e^2$ in powers of $\theta$.
Eliminating $\theta$ then leads to the PC expansion of the gauge-invariant relation between $\g$ and $J/e^2$. In terms
of the notation used in the previous sections, i.e. writing $\frac{e^2}{J} = - \frac{A}{j}$, we found
\bea \label{gvsjnu}
\gamma^{\rm circ, Schild}(j)&=&1- \frac12 \frac{A^2}{j^2}   -\frac18 \frac{A^4}{j^4}  + \frac{A^6}{j^6} \left(- \frac{1}{16}  + \frac{\nu}{2}\right)\nonumber\\
&+& \frac{A^8}{j^8} \left(- \frac{5}{128}  +  \frac{5\nu}{18}\right)\nonumber\\  
&+&\frac{A^{10}}{j^{10}} \left(- \frac{7}{256 } +  \frac{11}{30} \nu   -  \frac{101}{72} \nu^2 \right)\nonumber\\
&+& + 
 \frac{A^{12}}{j^{12}} \left(- \frac{21}{1024}  +  \frac{556}{1575} \nu   -  \frac{41}{30} \nu^2   + \frac12 \nu^3 \right)\nonumber\\ 
&+& \frac{A^{14}}{j^{14}} \left(- \frac{33}{2048}  +  \frac{6023}{17010} \nu   -  \frac{4590857}{1814400} \nu^2  \right.\nonumber\\
& +&\left.  \frac{1793}{288} \nu^3  \right)\nonumber\\
& +& \frac{A^{16}}{j^{16}} \left(- \frac{429}{32768}  +  \frac{772841}{2182950} \nu   -  \frac{3398741}{1088640} \nu^2\right.\nonumber\\ 
  &+&\left.  \frac{24035}{2592} \nu^3   -  \frac{101}{18} \nu^4   \right)\nonumber\\ 
&+& \frac{A^{18}}{j^{18}} \left(- \frac{715}{65536}  +  \frac{2152924}{6081075} \nu  -  \frac{896088883}{223534080} \nu^2\right.\nonumber\\   
&+&\left.  \frac{23785303}{1209600} \nu^3  
-  \frac{1885697}{51840} \nu^4   +  \frac{3}{2} \nu^5 \right)\nonumber\\
&+& O\left(\frac{A^{20}}{j^{20}}\right)\,.
\eea
In the probe limit ($\nu \to 0$) this  reduces (in keeping with Eq. \eq{gvsjprobe}) to the large-$j$ expansion of $ \sqrt{1 - \frac{A^2}{j^2}}$, while in the equal mass case it yields  
\bea
\gamma|^{\rm circ, Schild}_{\nu=\frac14}(j)&=&1 -\frac{A^2}{2j^2} -\frac{A^4}{8j^4}+\frac{A^6}{16j^6}\nonumber\\
&+&\frac{35
   }{1152}\frac{A^8}{j^8}-\frac{269 }{11520}\frac{A^{10}}{j^{10}}-\frac{15899}{1612800} \frac{A^{12}}{j^{12}}\nonumber\\
&+&\frac{1005469
   }{87091200}\frac{A^{14}}{j^{14}}
+\frac{349770523 }{107296358400}\frac{A^{16}}{j^{16}}\nonumber\\
&-&\frac{3529853293}{557941063680} \frac{A^{18}}{j^{18}}\nonumber\\
&+&O\left(\frac{A^{20}}{j^{20}}\right)\,.
\eea

\section{Using Schild's exact solution to derive constraints beyond  $\alpha^5$.}

In this final Section, we use the high-order expansion of the gauge-invariant relation $\gamma^{\rm circ, Schild}(j)$, Eq. \eq{gvsjnu},
to:
(i) obtain checks of the  $O(\alpha^5)$ scattering angle derived in Ref. \cite{Bern:2023ccb};
and (ii) derive constraints on the dynamics of electromagnetically interacting binary systems beyond  $O(\alpha^5)$.
As already mentioned, we use as go-between the EOB description of the (conservative) binary dynamics via
the energy-dependent electric potential $\phi(\g,\nu,u)$.

From the $O(\alpha^5)$ scattering angle derived in Ref. \cite{Bern:2023ccb} we have derived above the
corresponding exact values of the PL-expansion coefficients, $\phi_n(\g, \nu)$, of the EOB potential encoding the binary dynamics.
Starting now from the knowledge of the $\phi_n(\g, \nu)$'s up to $n=5$, we can use the circular-motion equations, Eqs. \eq{ju} and \eq{gu}, 
which give, by eliminating the auxiliary variable $u$ between them, the relation between $j^2$ and $\g$. As explained above,
one can perturbatively eliminate $u$ so as to get  $\g_{\rm circ}-1$ as a power series in $\frac{A^2}{j^2}$.
Eq. \eq{gvsy} above illustrates that the coefficient of $\frac{A^{2n}}{j^{2n}}$ involves the PL-expansion coefficients of
the potential up to  $\phi_n$.  The relation between energy and angular momentum derived from
the $O(\alpha^5)$-accurate knowledge of  $\phi(\g, \nu, u)$ is then obtained as a PC expansion whose accuracy is correspondingly
limited to the fourth PC-order, i.e. a factor $\frac{A^{8}}{j^{8}}$ beyond the leading Coulomb circular energy 
$\g_{\rm circ}-1 \approx -\frac{A^2}{2j^2}$, i.e. it yields the knowledge of the $\nu$-dependent coefficients
$C_4(\nu)$, \ldots,  $C_{10}(\nu)$, in the following limited expansion:
\bea
\g_{\rm circ}^{ \alpha^5 \rm scattering}-1 &=& -\frac{A^2}{2j^2} + C_4(\nu) \frac{A^4}{j^4}+ \cdots \nonumber\\ 
&+&   C_{10}(\nu) \frac{A^{10}}{j^{10}} + O\left(  \frac{A^{12}}{j^{12}}\right)\,.
\eea
We have verified that all the coefficients $C_4(\nu)$, \ldots,  $C_{10}(\nu)$ derived from the $O(\alpha^5)$ scattering angle of \cite{Bern:2023ccb},
via the correspondingly accurate EOB potential, agree with the corresponding coefficients in Eq. \eq{gvsjnu}, which we derived
from the exact two-body solution of Schild \cite{Schild:1963}. Note in passing that this check being done at the 4PC order 
has, in fact, only used the $\pinf$ expansions of the various potential coefficients $\phi_2(\g,\nu), \cdots, \phi_5(\g,\nu)$
at  a finite order (namely two $\pinf^2$ orders less than those indicated in Eqs. \eq{phinPCexp}, which are needed to go to the 6PC order).

After this check, we can now go beyond the $O(\alpha^5)$ accuracy by using the knowledge of $\gamma^{\rm circ, Schild}(j)$
at the 8PC order displayed in Eq. \eq{gvsjnu} above. For instance, if we work at the combined 7PL and 6PC orders, i.e.
if we start from the general parametrization of $\phi_6$ (at $\pinf^0 + \pinf^2$-accuracy) and $\phi_7$ (at $\pinf^0$-accuracy)
given in Eqs. \eq{phi67PCexp}, the expansion, Eq. \eq{gvsy}, yields for $\g^{\rm circ}-1$ an expansion up to 
the 7PL+6PC order, i.e. up to  $\frac{A^{14}}{j^{14}}$, where the coefficient of  $\frac{A^{12}}{j^{12}}$
involves $\phi_6^{(0)}$, while the coefficient of  $\frac{A^{14}}{j^{14}}$ involves a linear combination of
$\phi_6^{(2)}$ and  $\phi_7^{(0)}$. By comparing this expansion with our Schild-derived expansion Eq. \eq{gvsjnu},
we could derive the following new results (belonging to the 6PL and 7PL orders):
\bea \label{phi60}
\hat \phi_6^{(0)}&=& \nu\left( - \frac{15751}{7200}    - \frac{ 331}{180}\nu   -  \frac{29}{72} \nu^2 
   -  \frac{9}{32}\nu^3 \right), 
  \eea
  and 
 \bea \label{phi762}
\hat \phi_7^{(0)}+\hat \phi_6^{(2)}&=&\nu\left(  \frac{230493299}{95256000}  + \frac{ 81533}{22680} \nu   +  
     \frac{2369}{720} \nu^2   \right.\nonumber\\
&+&\left.  \frac{683}{576} \nu^3  +  \frac{3}{4} \nu^4\right)\,. 
\eea
Note that our result Eq. \eq{gvsjnu} gives access to higher PL levels, namely up to the 9PL level, corresponding to
$\alpha^9$. However, because of our limitation to circular orbits, we do not have access to the full functions
$\hat \phi_n(\pinf^2)= \sum_{k\geq0} \hat \phi_n^{(2k)} p_\infty^{2k}$, but only to combinations of  
various coefficients  $ \hat \phi_{n_1}^{(2k_1)}$ belonging to the same PC order.

Having in hands, the new knowledge, Eqs. \eq{phi60}, \eq{phi762}, about  6PL and 7PL orders, we can then use the EOB-potential
computation of the scattering angle to obtain corresponding information about the scattering angle at 6PL and 7PL orders.
More precisely, using Eqs. \eq{chinvsphin}, we can compute the 6PL and 7PL scattering coefficients $\chi_6$ and
$\chi_7$ at  the 6PC order. Let us work for simplicity with the energy-rescaled (and $A$-rescaled)  scattering coefficients, say
\be
\check \chi_n \equiv \frac{h^{n-1} \chi_n }{A^n} . 
\ee
We recall the values of the probe limits of $\check \chi_6$ and $\check \chi_7$:
\bea
\check  \chi_{6,0} &\equiv &\check \chi_6 (\nu=0) = \frac{5\pi}{32}\,,\nn
\check  \chi_{7,0} &=&\check \chi_7(\nu=0) = -\frac{\gamma(16\gamma^6 - 56\gamma^4 + 70\gamma^2 - 35)}{35(\gamma^2-1)^{7/2}} \,.\nonumber\\
\eea
Then, our Eqs. \eq{chinvsphin} yield: 
\bea \label{chi6phi}
\frac{\check  \chi_{6}  - \check  \chi_{6,0}}{\pi }  &=&\frac{5\nu}{96}(-29+45\nu) +\frac{\nu}{64}(-82+145\nu)\pinf^2\nonumber\\  
&+& \left(\frac{5855}{10752} \nu   -  \frac{15}{16} \hat\phi_{6}^{(0)}  -  \frac{53}{128} \nu^2\right.\nonumber\\ 
&-&\left.  
\frac{145}{384} \nu^3  -  \frac{135}{512} \nu^4  \right) \pinf^4\nonumber\\
&+& \left(\frac{2933621}{11289600} \nu  - \frac{15}{16} \hat\phi_{6}^{(2)}  - \frac{326761}{107520} \nu^2\right.\nonumber\\  
&-&  
 \frac{499}{3072} \nu^3  - \frac{365}{1536} \nu^4  + \frac{105}{512} \nu^5\nonumber\\  
&-&\left. 
\frac{15}{32}\hat\phi_{6}^{(0)} \left(1  +5\nu \right)  \right)\pinf^6\,,
\eea
and 
\bea \label{chi7phi}
\check  \chi_{7}  - \check  \chi_{7,0}  &=&\frac{3  \nu}{7 \pinf^5} 
+ \frac{ 3 \nu (-15 + 4 \nu) }{28 \pinf^3 } \nonumber\\
&+& \frac{\nu (1777 - 3888 \nu + 72 \nu^2)}{ 504 \pinf}\nonumber\\ 
&+& \frac{\nu (64273 - 161780 \nu + 22800 \nu^2) }{ 6720} \pinf\nonumber\\ 
&+& \left(\frac{3799207}{201600}\nu + 16 \hat \phi_6^{(0)} + \frac{879}{80 } \nu^2\right.\nonumber\\ 
&+&\left.  
 \frac{3611}{336} \nu^3  + \frac{9}{2} \nu^4  \right)  \pinf^3\nonumber\\ 
&+&\left( -\frac{2691959153}{84672000}\nu +  \frac{96}{5} \hat \phi_6^{(0)}  + 16\hat \phi_6^{(2)}\right.\nonumber\\ 
&-&  
    \frac{16}{5} \hat \phi_7^{(0)}   +  \frac{217743}{4480}\nu^2 + 48 \hat \phi_6^{(0)}\nu\nonumber\\   
&+&\left.  
 \frac{886}{45} \nu^3  + \frac{409}{36} \nu^4  - \frac{9}{5} \nu^5 \right)\pinf^5\,.
\eea
As we see on these expressions the $O(\pinf^4)$ contribution to $\check  \chi_{6} $, as well as the  $O(\pinf^3)$ contribution to $\check  \chi_{7} $ only depend on $\hat \phi_6^{(0)}$. Therefore, the value, Eq. \eq{phi60}, that we derived from Schild's solution
allows us to predict the value of the $O(\al^6)$ contribution to the scattering angle through the $\pinf^4$ accuracy, namely
\bea \label{chi6p4}
\frac{\check  \chi_{6}  - \check  \chi_{6,0}}{\pi }  &=& \frac{5\nu}{96}(-29+45\nu) +\frac{\nu}{64}(-82+145\nu)\pinf^2\nonumber\\ 
&+& \frac{\nu}{13440}(34883+17605\nu)\pinf^4\nonumber\\
&+&  O(\pinf^6)\,.
\eea
As expected, the $O(\nu^3)$ and $O(\nu^4)$ terms that were present in  $\hat \phi_6^{(0)}$,  Eq. \eq{phi60}, have
cancelled against contributions coming from $\phi_1, \cdots, \phi_5$ to predict a $\pinf^4$-accurate value of $\check  \chi_{6} \equiv \frac{h^{5} \chi_n }{A^6}$ which contains only $O(\nu)$ and $O(\nu^2)$ contributions.

Considering now the $O(\pinf^6)$ contribution to $\check  \chi_{6} $, which depends on $\hat \phi_6^{(2)}$, we cannot
predict its full value, because the Schild circular solution only gave us access to the sum $\hat \phi_6^{(2)}+ \hat \phi_7^{(0)}$.
Let us write $\hat \phi_6^{(2)}$ in the general form,
\be
\hat \phi_6^{(2)}= b_1 \nu + b_2 \nu^2 + b_3 \nu^3+  b_4 \nu^4 + b_5 \nu^5.
\ee
Inserting such a parametrized value in the $O(\pinf^6)$ contribution to $\check  \chi_{6} $ in Eq. \eq{chi6phi} yields a
coefficient which contain powers of $\nu$ up to $\nu^5$. However, we know that $\check  \chi_{6} $ should be (at all orders
in $\pinf$) a quadratic polynomial in $\nu$. The latter condition gives us three equations determining the values
of the coefficients $b_3$, $b_4$ and $b_5$ entering $\hat \phi_6^{(2)}$. Namely, we get
\bea
b_3&=& \frac{13321}{2880}\,,\nonumber\\   
b_4&=& \frac{515}{576}\,, \nonumber\\    
b_5&=&\frac{59}{64}   \,.
\eea
In other words, we cannot predict from our current results the $O(\pinf^6)$ contribution to $\check  \chi_{6} $, but we can
predict most of the $\nu$-structure of the potential coefficient $\hat \phi_6^{(2)}$, namely
\be \label{phi62final}
\hat \phi_6^{(2)}= b_1 \nu + b_2 \nu^2 + \frac{13321}{2880}\nu^3+ \frac{515}{576}\nu^4 +\frac{59}{64} \nu^5\,.
\ee
Inserting these values in Eq. \eq{chi6phi} finally yields the following expression for $\chi_6$  through order $\pinf^6$
\bea \label{chi6p6}
&&\frac{\check  \chi_{6}  - \check  \chi_{6,0}}{\pi } =\frac{5\nu}{96}(-29+45\nu) +\frac{\nu}{64}(-82+145\nu)\pinf^2\nonumber\\ 
&&\qquad +\frac{\nu}{13440}(34883+17605\nu)\pinf^4\nonumber\\ 
&&\qquad +\left[-\frac{15}{16}(b_1 \nu+ b_2 \nu^2)+\frac{7255303}{5644800} \nu+\frac{79301}{26880}\nu^2  \right]\pinf^6\nonumber\\ 
&&\qquad +O(\pinf^8)\,.
\eea
Furthermore, inserting the result \eq{phi62final} in our constraint, Eq. \eq{phi762}, we find the following value for the
7PL potential coefficient $\hat \phi_7^{(0)}$:
\bea \label{phi70}
\hat \phi_7^{(0)}&=&\left (\frac{230493299}{95256000}- b_1 \right)\nu + \left(\frac{81533}{22680}-b_2 \right)\nu^2\nonumber\\ 
&-&\frac{769}{576} \nu^3 + \frac{7}{24} \nu^4 -\frac{11}{64}\nu^5\,.
\eea
Finally, inserting the results so obtained for: $\hat \phi_6^{(0)}$, Eq. \eq{phi60}, $\hat \phi_6^{(2)}$, Eq. \eq{phi62final},
and $\hat \phi_7^{(0)}$, Eq. \eq{phi70}, in Eq. \eq{chi7phi}, we get the following prediction for $\check  \chi_{7} $:
\bea \label{chi7final}
\check  \chi_{7}  - \check  \chi_{7,0}  &=& \frac{3  \nu}{7 \pinf^5} 
+ \frac{ 3 \nu (-15 + 4 \nu) }{28 \pinf^3 } \nonumber\\
&+& \frac{\nu (1777 - 3888 \nu + 72 \nu^2)}{ 504 \pinf}\nonumber\\ 
&+& \frac{\nu (64273 - 161780 \nu + 22800 \nu^2) }{ 6720} \pinf\nonumber\\
&+& \frac{\nu (-3257241 - 3716440 \nu + 867400 \nu^2)}{201600}  \pinf^3\nonumber\\
&+& \left[\frac{96}{5} (b_1\nu+ b_2 \nu^2)-\frac{310681544797}{3810240000} \nu \right.\nonumber\\
&-&\left.\frac{37454209}{362880} \nu^2+\frac{59}{30}\nu^3 \right] \pinf^5 \nonumber\\
&+&O(\pinf^7) \,.
\eea
Note that the same combination $b_1 \nu+ b_2 \nu^2$ (inherited from $\hat \phi_6^{(2)}$)
enters both Eqs. \eqref{chi6p6} and \eqref{chi7final}.

In other words, we predict the value of $\check  \chi_{7}$ through order $O(\pinf^3)$, and we also predict the $O(\nu^3)$
contribution to $\check  \chi_{7}$ at order $O(\pinf^5)$. [Note that the insertion of our value of  $\hat \phi_6^{(0)}$ in
the $O(\pinf^3)$ contribution to $\check  \chi_{7}$ has led to  the cancellations of high powers of $\nu$ needed
to end up with a result cubic in $\nu$.]
When the value of $\chi_6(\pinf)$ is derived, our result \eq{chi7final} will give a precise prediction for $\chi_7$ through $O(\pinf^5)$. 

\section{Conclusions}

In the present work we have shown how the electromagnetic version of the EOB formalism allowed one to
transfer information back and forth between the post-Lorentzian expansion of the conservative scattering function,
$\chi(E,J)$, and the post-Coulombian expansion of the energetics of bound states. 
We have provided checks of the $O(\al^5)$ scattering results of Bern et al. \cite{Bern:2023ccb}.
Using the classic circular electrodynamical solution of Schild \cite{Schild:1963},
we have derived predictions about the conservative  scattering angle at orders  $O(\al^6)$ and  $O(\al^7)$,
namely, Eqs. \eq{chi6p4}, \eq{chi6p6} and \eq{chi7final}. These predictions give benchmarks for future,
higher-order computations of the conservative scattering of electromagnetically interacting binary systems.
It would also be interesting to explore whether the square-root vanishing of $E^{\rm circ}(J, \nu)$ at the
universal value $J_m/e^2= {\mathcal J}_m = 0.4416107917053\ldots$, Eqs. \eq{Jm}, \eq{sqrtvanishing} , reflects itself in the behavior of 
the scattering function $\chi(\g, \nu, J/(e_1e_2))$. [Note that $E=0$ corresponds to the negative value $\g= 1-\frac1{2\nu}$,
and $J/e^2$ to $- J/(e_1e_2)$.]

Let us mention some possible extensions of our work. First, by perturbing the exact circular solution of Schild, one might
be able to derive a small-eccentricity solution of an electromagnetically interacting binary system; the
corresponding knowledge of the periastron precession, $K^{\rm circ}(J)$,  would give us
the $O(I_r^{\rm ell})$ contribution to the Delaunay Hamiltonian, $E = H(I_r^{\rm ell}, J)$, say 
\be
 H(I_r^{\rm ell}, J)
= H^{\rm circ}(J) +\frac{d H^{\rm circ}(J)/dJ}{ K^{\rm circ}(J)} I_r^{\rm ell}+ O\left(( I_r^{\rm ell})^2\right).
\ee
In turn, this  new contribution would give (via the EOB potential $\phi$)  additional information about the
scattering function beyond $\al^5$.

Second, one might think of perturbatively including radiation-reaction effects in Schild's conservative solution, thereby
acquiring some knowledge about corresponding radiative effects in the scattering function.
Finally, though there are no exact analytical gravitational analog of Schild's circular binary solution, the helical-Killing-vector
numerical solutions of Einstein's equations \cite{Detweiler:1989,Gourgoulhon:2001ec,Grandclement:2001ed} might provide interesting information 
about the conservative dynamics of gravitationally interacting binary systems.

\section*{Acknowledgements}
D.B. 
acknowledges sponsorship of the Italian Gruppo Nazionale per la Fisica Matematica (GNFM)
of the Istituto Nazionale di Alta Matematica (INDAM), 
as well as the hospitality and the highly stimulating environment of the Institut des Hautes Etudes Scientifiques.
The present research was partially supported by the 2021
Balzan Prize for Gravitation: Physical and Astrophysical
Aspects, awarded to T. Damour.

 \end{document}